# A FRAMEWORK FOR EXTRACTING AND MODELING HIPAA PRIVACY RULES FOR HEALTHCARE APPLICATIONS

Tariq Alshugran and Julius Dichter

Department of Computer Science and Engineering, University of Bridgeport,
Bridgeport, CT, USA
talshugr@my.bridgeport.edu, dichter@bridgeport.edu

## ABSTRACT

*Some organizations use software applications to manage their customers' personal, medical, or financial information. In the United States, those software applications are obligated to preserve users' privacy and to comply with the United States federal privacy laws and regulations. To formally guarantee compliance with those regulations, it is essential to extract and model the privacy rules from the text of the law using a formal framework. In this work we propose a goal-oriented framework for modeling and extracting the privacy requirements from regulatory text using natural language processing techniques.*

## 1. INTRODUCTION

Software applications are developed to help companies and organizations to process and manage data that support their daily operations. However, this data might contain sensitive clients' information that should be protected to ensure clients' privacy. Besides losing clients' trust, neglecting to ensure the clients' data privacy may also be unlawful and inflict serious legal and financial consequences. Lately, different laws and regulations [1]–[3] related to data privacy have been enacted specially in vital sectors such as health care, finance, and accounting. Those regulations dictate how clients' data should be disclosed and transmitted within the organization and also with external partners. The privacy rules in laws and regulations presented a challenge for software engineers who design and implement software applications that process private client data. The difficulty is linked to the complexity and length of the letter of the law and the how to guarantee that the software application is maintaining the clients' data privacy in compliance with the law

Some healthcare organization are trying to perform their own interpretation of the law privacy rules by creating custom systems. However, the problems with such approach is that the margin of error while interpreting the letter of the law is high specially with separate efforts carried out by individual companies. According to a survey carried out to check the Healthcare Insurance Portability and Accountability Act (HIPAA) requirements interpretation created for medical and healthcare related applications, none of the frameworks were well developed to capture the relationships specified in the law [4]. To solve this problem, a standard framework is required that will analyze the regulatory text and provide a method to extract the relevant component that can be used during software roles engineering and development. The extracted components will include all the possible arrangements of roles, purposes, permissions, temporal factors, and any carried out obligations.

In this work we propose a framework to analyzes, extracts, and models the privacy requirements from HIPAA regulatory text. The framework goal is to translate the law privacy rules text into more manageable components in the form of entities, roles, purposes, and obligations. Those components together can be used as building blocks to create formal privacy policies. The process concentrate on two main components; entities and their roles, and data access context. To get the first part, the framework will parse the privacy sections of the regulatory text to mine all the subjects, and then categorize those subjects into roles based on their characterization in the law. To acquire the access context, the process will extract all the purposes, temporal clauses and any carried out obligations and classify them based on their permissibility.

The rest of this work is organized as follows. Section 2 examines the complexity of HIPAA and other regulatory text modeling. In section 3, we cover access control models and the elements that should be extracted based on the selected access control mode. Whereas section 4 describes the proposed framework for extracting and modeling the access requires context from the regulatory text. Section 5 provides a literature survey of laws and regulations modeling and current proposed approaches. Finally, we discuss our future work and conclude in section 6.

## 2. LAWS' PRIVACY RULES AND MODELING COMPLEXITY

In the U.S., numerous federal laws and regulations were legislated to guarantee individuals' right to be able to access and port their private information stored and managed by service providers while protecting that information from unauthorized access. For example, the Gramm-Leach-Bliley Act (GLBA) of 1999 [2] is designed to protect individuals' financial information from being breached without proper authorization. The Health Insurance Portability and Accountability Act (HIPAA) of 1996 [1] is another example, section 164 of HIPAA is intended explicitly to protect patients' healthcare information and medical records from unauthorized disclosure [5]. According to HIPAA, any healthcare related information that can identify an individual and can be stored or transmitted via any media format is defined as Protected Health Information (PHI). HIPAA privacy rules control the storage, transmission, and disclosure of all Protected Health Information. Usually the PHI is collected and maintained by healthcare insurance plan, healthcare provider, healthcare clearinghouse or any other similar organization identified by HIPAA as Covered Entity. Comparable privacy rules can also be found in other federal regulations as well.

### 2.1. Modeling Complexity

The U.S. federal regulations documents are written in a complex format and technical language known as legalese. Legalese or legal English uses different vocabulary and syntax than that used in ordinary English. The complex format and legal terminologies makes legal documents hard to read and interpret.

A part from the document language, the structured format makes the text prone to misinterpretations and other ambiguities like cross references and exceptions. The document is usually structured into parts (e.g. Part 164 of HIPAA). Each part is then divided into subparts, which is additionally divided into sections (e.g. Section §164.528). Some sections are also divided into subparagraphs with multiple points in the same sentence. This create some inconsistency as some privacy rules are spanning multiple points, subparagraphs, paragraphs, or event sections. For example, the subparagraph §164.528(a)(2)(ii) contains three points (A), (B), and (C) in the same sentence: "the covered entity must:(A)...;(B)...; and (C)...", where each one of these points defines a different obligation that should be carried out by the covered entity.

Exceptions and Cross-references to other sections add more complexity to the modeling process as they require an additional processing efforts. References usually entail priority between paragraphs and add more clarity to the privacy requirements. However, sometimes references might introduce ambiguities due to the possibility of nested and multilevel referencing in the form of cross-references. Cross-references occur when a section or a paragraph in the law is referencing another section/paragraph that has a reference to another rule. For example, the subparagraph §164.528 (a)(2)(i) describes individuals' suspension of rights obligation. This right is also addressed in a different paragraph as highlighted by the phrase "as provided in §164.512(d)" at the end of the paragraph establishing a reference. Nevertheless, in the subsequent paragraph §164.528(a)(2)(ii), the phrase "pursuant to paragraph (i)" is a reference to the preceding paragraph. Hence, to model the right indicated in §164.528(a)(2)(ii), we need to refer to §164.512(d) creating an indirect relation between the two paragraphs. On the other hand, exceptions are rules that contradict or negate other rules by changing the permissibility right or by adding more conditions or obligation. So, if the first rule grants a right to access a PHI, the exception would either add more conditions to clarify that right or grant permission and vice versa. For instance, §164.512(c)(1) in HIPAA grants the right to disclose a PHI if the information is about adult victims of abuse, neglect or domestic violence, however, §164.512(c)(1)(ii) presents an exception to this right by adding victims agreement as a condition for such a disclosure.

## 3. ACCESS CONTEXT

Access context represents the elements that can be used as an input to the to an access control system. To create a privacy preserving access control system that enforces the privacy rules from the regulatory text, we need to extract the access context. In this section we show those components and the importance of formalizing them.

### 3.1. Context-Based Access Control

In information security, Access Control is implemented as a mean to decide whether any specific authenticated system user has the proper permission to access a certain data object, or carry out a particular type of operation (e.g. read, write, delete …etc.) on that object [6].

Most access control models are described using three terms: subject, object, and operation. Subject refers to any authenticated user, it could be a system, a person, or a process. Whereas object is defined as the private data that the access control system is protecting. Finally, operation is any action that can be taken by the subject on the object. Figure 1 shows the three terms of simple access control model. The permissibility to carry out those operations is ruled by a collection of access rights expressed in the format:

ALLOW [*Subject*]
TO PERFORM [*Operation*] ON [*Object*]

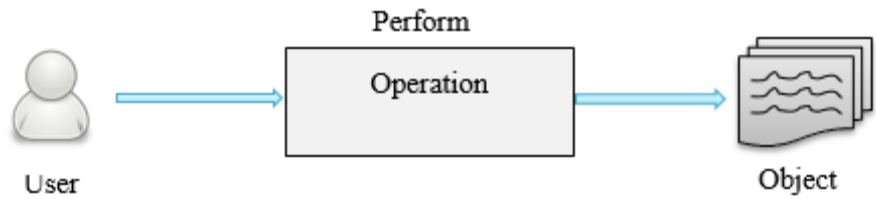

Figure 1. Simple Access Control Model

Access control models evolved over time to solve particular issues the traditional model was not able to handle correctly. For instance, Role-Based Access Control model (RBAC) [7] extended the traditional simple access control model described above to add the subject role. In RBAC users' role is used in the access right rules instead of users' identities. RBAC model was introduced to solve the any inconsistency faced by dynamic systems where users' role is prone to change, and hence, there permission and access rights should change as well to adapt the new roles. Figure 2 below illustrate RBAC components and the relation with the system users. RBAC access rights rules are formally formatted and expressed as:

ALLOW [*Role*]
TO PERFORM [*Operation*] ON [*Object*]

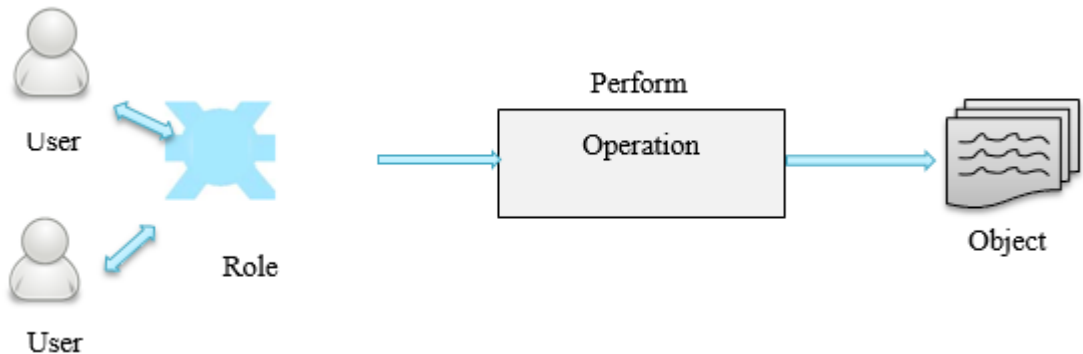

Figure 2. Role-Based Access Control (RBAC) Model

In order to facilitate a more fine-grained details that will capture the privacy rules and requirements of HIPAA regulatory text, a new model was introduced to capture the access context. The Context-Based Access Control (CBAC) model is an extension to RBAC with more details to allow higher flexibility and regulatory compliance. CBAC depends on the user roles, request purpose, and object type to determine the permissibility of the access request. CBAC also provides a mechanism to log any carried out obligations as a result of granting access to the object. Recently, many formats were suggested to represent CBAC, in this work we propose a modified version Powers et al [6] version of the Privacy Access Control model as shown in figure 3. CBAC access rules can be formally expressed in the format below:

ALLOW [*Active Role*]
TO PERFORM [*Operation*]
ON [*Data Type*]
RELATED TO [*Data Owner Type*]
FOR [*Purpose*]
PROVIDED [*Condition*]
CARRY OUT [*Obligation*]

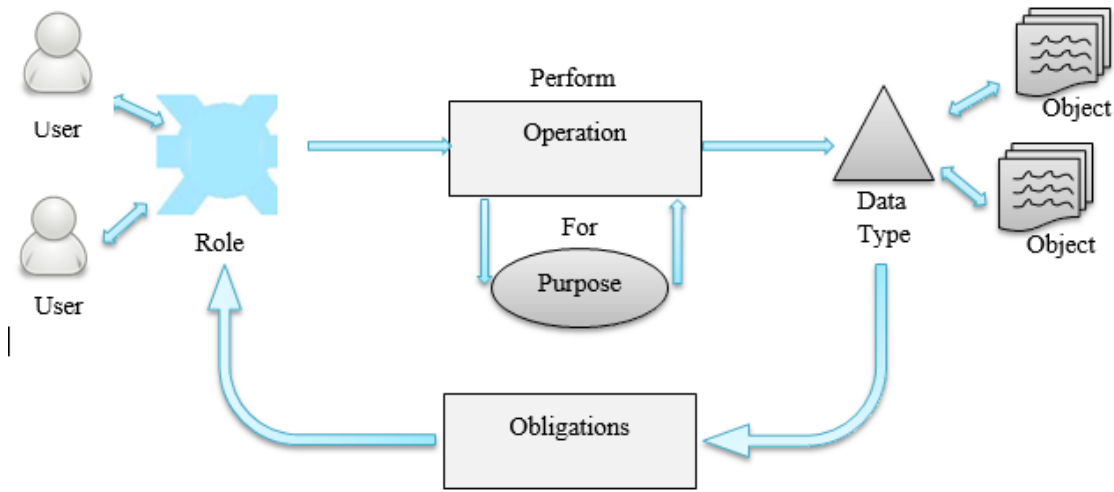

Figure 3. Context-Based Access Control (CBAC) Model

**3.2. CBAC Components**

To better understand what elements and components the framework should extract from regulatory text, we need to align the extraction and modeling process with CBAC access rules. In this section we will describe each component and how to formally model that component in order to extract it.

*Requester Active Role:* As CBAC is an extension of RBAC, it also uses the requester role instead of the requester identity. However, due to the fact that any user can be in multiple roles, we use the current active role only. For example, if Alice is a doctor and a patient at the same time, when accessing her own medical file she will be treated as a patient only as that is the current active role of Alice.

*Operation:* any set of actions that can be applied to data objects, similar to the traditional model. The set contains read, write, and delete action in addition to any other applicable operation provided by the system.

*Data Type:* Context-Based access rules are not tied directly to a particular data object, but rather to the type of data object. The data type can be defined at high-level (e.g. treatment related files), or at a much lower-level (e.g. Medical chart files, CT scan images …etc.). This introduces the notion of attaching attributes or metadata to the data itself.

*Data Owner Type:* The data owner type element specifies the state and the type of the owner of the data object. By capturing this element, we can establish a relation between the PHI, its owner, and the requester. For example, if Alice is a doctor requesting access to Bob's medical file, we can check if Alice is role as a doctor is relevant to Bob's file as a patient of Alice. If the relation is established then in this case Bob is the owner of the PHI and his type in relation to Alice is a patient.

*Purpose:* This element represents the reason why the access requester is requesting access to the specified data object type. Purposes can also be classified at a high-level or low-level depending on the provided request context and the letter of the law. It could be healthcare related like treatment or medical consulting, financial for the purpose of accounting, or legal like violence and crime investigations.

*Conditions:* Any pre-defined additional conditions and criteria required to be fulfilled *before* accessing the protected data object or allowing the disclosure. A common example of conditions in HIPAA is obtaining the data owner permission and consent before sharing their data with any third party.

*Obligations:* Obligations in access request refer to the action that should be carried out by the covered entity, the access requester, or the system itself *after* permission was granted and the data was transmitted. It might include operations like logging for audit purposes or notifying the data owner of the disclosure action.

## 4. THE MODELING PROCESS

This section presents the proposed framework process and steps used to extract privacy requirements from the regulatory text. The process is divided into two activities: firstly is the model and analysis activities to model the regulatory text, and secondly the identification and extraction activity to extract the privacy requirements context. Each one of the two activities contains a number of steps. Figure 4 demonstrates the extraction process with the two activities as well as each activity initial inputs and the expected outputs.

### 4.1. Model and Analyze HIPAA

The purpose of this activity is to overcome the difficulties and issues that complicate HIPAA modeling as explained in section II.B. Raw HIPAA rules are used as an input to the activity, and then the following steps are performed:

***Step 1:*** Identify scopes and definitions: the first step is to outline the extraction scope. In this framework, the emphasis will be on the privacy rules related to preserving patients' identity and PHI disclosure. After analyzing HIPAA regulatory text and excluding all the abstract, non-technical requirements, we found that the extraction process should only be applied to the privacy requirements from the following subsections of HIPAA: §164.502, §164.506, §164.510, §164.512, §164.514, and §164.524.

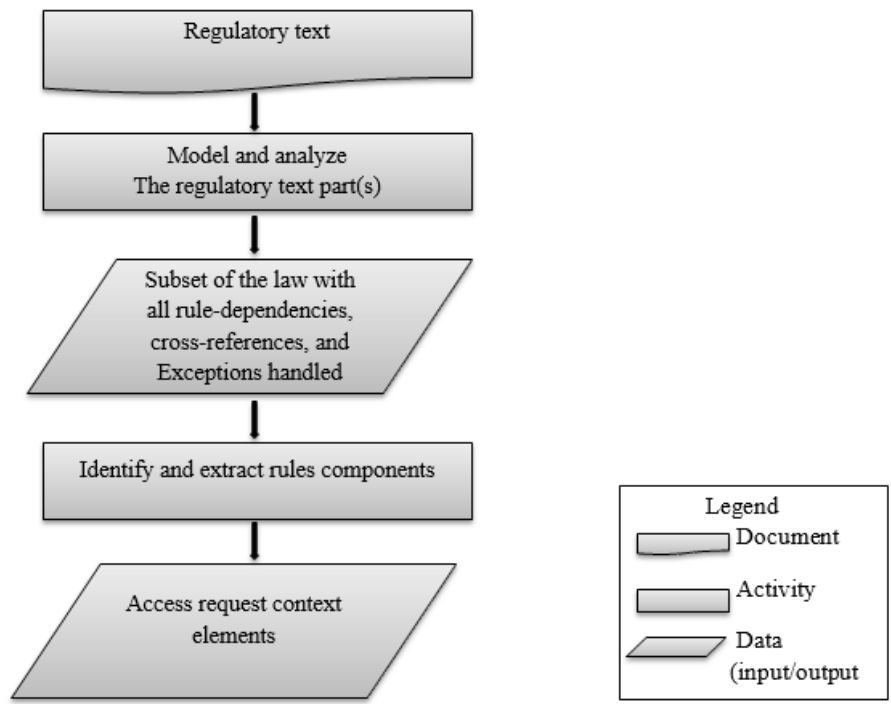

Figure 4. HIPAA privacy requirements extraction process

***Step 2:*** Identify and resolve rules dependencies, cross-references, and exceptions: the goal of this step is to validate the subset of selected rules from step 1 and to add more clarity to HIPAA legal text. This goal is attained by replacing self-references, dependencies, and cross-references, with the exact description from the referenced rule, or by applying a similar depiction to improve the rule readability and interpretation. Alternatively, if the reference is identified as a condition it might be replaced with a rule identifier placeholder for further processing in an advanced stage of the process.

The output of the first activity is a reference-free less-ambiguous subset of HIPAA rules that is focused only on patients' privacy. This output will then be used as an input in the following activity.

## 4.2. Identify and Extract the Context Elements

The second activity aims to perform the extraction process. Similar to the preceding activity, this activity also is divided into multiple steps where the final step is the responsible for the context elements extraction. In this activity the regulatory text is parsed using a Natural Language Processing (NLP) application to highlight phrases that may establish a possible context element. Figure 5 outlines the components of the rule §164.528 (a)(2)(i) of HIPAA as an example of the expected results.

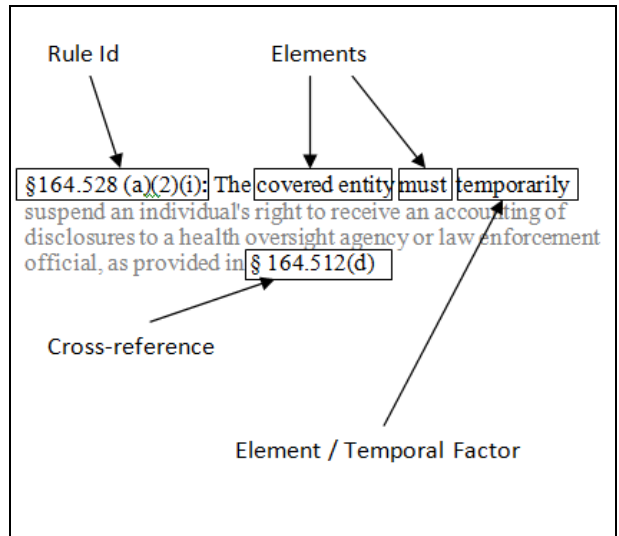

Figure 5. An outline of HIPAA rule §164.528 (a)(2)(i) components.

***Step 1***: Clearing ambiguities: HIPAA legal text contains domain-specific wording and terminologies that will require the assistance of law professional to provide a precise interpretation and definition of the text. As some words, terms, and phrases might carry multiple meanings causing further ambiguity and a higher chance of misinterpretation. At this step, all the ambiguous terms, words, and phrases will be mapped to a set of possible meanings as proposed by the work of Otto et al [8].

***Step 2***: Text parsing and elements extraction: a text parser is used to identify any possible context element. Rules of generalization and specialization are also applied to specify generic roles.

Table 1. Subparagraph §164.528 (a)(2)(i) Elements Classification

| *Element* | *Element Classification* |
|---|---|
| R164_528_a_2_i | Rule Id |
| covered entity | Role |
| must | Operator |
| temporarily | Temporal factor |
| suspend rights to receive an accounting of disclosure | Action |
| health oversight agency | Role |
| law enforcement official | Role |
| (R164_512_d) OR (R164_512_d) | Condition |

***Step 3***: Elements Classification: Numerous approaches were proposed to classify elements [8]–[12]. Nevertheless, each approach was proposed to solve a particular problem or introduced a special notation that might not be applicable for generic access control. For instance, Hohfeld classification presented the notion of rights and responsibilities. Hohfeld theory represents the relationships between actors and the law based on their responsibilities and rights within the legal text context [13]. However, our framework is more concentrated on the concept of

Context-Based Access Control (CBAC) where the core focus is to identify roles, data type, purposes, conditions, temporal factors, and obligations. Hence, a Goal-Driven approach can be applied where the goal is to extract the CBAC components. Table 1 shows an example of elements classification constructed using a goal-driven approach on the data from §164.528 (a)(2)(i) of HIPAA.

## 5. RELATED WORK

Recent researches in the area of role extraction and engineering reveals a variety of proposed methods for extracting and modeling regulation components for different purposes. One of the oldest recognized efforts to classify the contents of regulatory text is attributed to Wesley Hohfeld which is known as Hohfeld legal taxonomy published in 1917 in Yale Law Journal [13]. Hohfeld taxonomy classifies regulatory text based on the notion of rights and obligations. Few recent researches have built their extraction models based on Hohfeld legal taxonomy like the work of Siena et al [14] and Islam et al [12]. Other approaches used Natural Language Processing (NLP) techniques to spot linguistic patterns to model the law and extract the context from the legal text [15]. In addition to the language patterns and Hohfeld taxonomy, other researchers used different methods including: analytical modeling based on organizational structure as in the work of Crook[9], UML-based [16], [11] method for the extraction of privacy requirements from organizations privacy requirements, and scenario-driven methods established for predefined set of tasks and scenarios [10]. Jorshari et al. work [17] focused on eliciting the security requirements in general without making the approach close enough to be adapted for extracting privacy requirements. Darimont et al [17] proposed a Goal-Oriented Requirements Engineering (GORE) approach to classify rules and form goals where those goals are then refined in an incremental fashion until all related tasks, actors, and uses-cases are discovered and extracted.

## 6. CONCLUSIONS

In this work, we proposed a framework to model the privacy requirements from regulatory text and to extract the possible context elements in the form roles, purposes, and obligations. The framework is designed to overcome the traditional complexities and challenges that face laws and regulations modeling. The framework consists of multiple steps starting by closely inspecting and analyzing the regulatory text to identify the parts of interest of the law that is related to the privacy requirements. Then it clarifies ambiguities from the letter of the law by resolving cross-references, dependencies, and handling rules exceptions. Next, a Goal-Driven approach is applied to examine the identified targeted rules text to extract all keywords that may define a context element.

In a later phase of this research, the context elements extracted using this framework will be used to create privacy policies as well as in the decision engine in an access control model. In order to implement such a model, the next phase will also include identifying the decision engine logic and the specification language that will be used for privacy policy representation.